\documentclass[preprint]{vgtc}               

\graphicspath{{figures/}{pictures/}{images/}{./}} 

\usepackage{times}                     

\usepackage{tabu}                      
\usepackage{booktabs}                  
\usepackage{lipsum}                    
\usepackage{mwe}                       

\usepackage{mathptmx}                  

\usepackage{longtable}
\usepackage[inline]{enumitem}
\usepackage{array}
\usepackage{fontawesome}

\newcommand{\sdno}{{\bfseries\color{BrickRed}\faClose}}
\newcommand{\sdyes}{{\bfseries\color{ForestGreen}\faCheck}}

\onlineid{0}

\vgtccategory{Research}

\vgtcinsertpkg

\title{
Bridging Service Design, Visualizations, and Visual Analytics in Healthcare Digital Twins: Challenges, Gaps, and Research Opportunities}

\author{Mariia Ershova\thanks{e-mail: mariia.ershova@uniroma1.it}\\ %
    \parbox{0.46\linewidth}{\scriptsize \centering Department of Planning, Design, Technology of Architecture \\ Sapienza University of Rome}
\and Graziano Blasilli\thanks{e-mail: graziano.blasilli@uniroma1.it}\\ %
     \parbox{0.47\linewidth}{\scriptsize \centering Department of Computer, Control, and Management Engineering \\ Sapienza University of Rome}}

\abstract{
    Digital twins (DT) are increasingly used in healthcare to model patients, processes, and physiological systems. While recent solutions leverage visualization, visual analytics, and user interaction, these systems rarely incorporate structured service design methodologies. Bridging service design with visual analytics and visualization can be valuable for the healthcare DT community. This paper aims to introduce the service design discipline to visualization researchers by framing this integration gap and suggesting research directions to enhance the real-world applicability of DT solutions.
} 


\begin{document}

\maketitle

\section{Introduction}
\label{sec:intro}

Digital Twins (\textbf{DT}s) are virtual representations and models of a physical system. 
Initially born in product lifecycle management, they have been used in different domains like the aviation sector, manufacturing, building and construction, and smart cities. DTs have recently increased in popularity in transforming healthcare through personalized virtual models, revolutionizing how patient care is delivered \cite{ATTARAN2023100165,Tianze2023,Tsekleves2025,Vallee2023}. 
%
Recent advances have pushed for the inclusion of visualization techniques \cite{Paziraei2025}, visual analytics (\textbf{VA}) \cite{Zheng2023}, and human-computer interaction (\textbf{HCI}) \cite{Barricelli2024} to the development of DT solutions. In healthcare, researchers have proposed solutions based on interactive dashboards, virtual reality interfaces, and immersive simulations that enhance the engagement of users (e.g., clinicians, caregivers, and patients) and facilitate decision-making \cite{Choi2025, Kim2024Assessing, Schmaltz2023, Xu2025Triboelectric}.
However, despite these technical achievements, users still have low involvement in the design and evaluation of such solutions \cite{Barricelli2024}. Most solutions focus on optimizing computational and visual aspects without considering the multi-actor (e.g., clinicians, caregivers, and patients) nature of healthcare services. As a result, many DT innovations risk poor integration into real-world workflows and limited adoption beyond proof-of-concept phases.
In parallel, \textit{service design} (\textbf{SD}) \cite{Stickdorn2012} has emerged as a robust approach for designing complex services that are user-centered, participatory, and systemically grounded. 
It is a multidisciplinary approach focused on designing and improving services by aligning user needs with organizational goals and system dynamics. SD aims to ensure that service is useful, usable, and desirable from the client’s perspective, as well as effective, efficient, and distinctive from the supplier’s perspective \cite{Stickdorn2012,Patricio2020}. Drawing from fields like interaction design, social design, and operations management, it follows principles such as user-centricity, co-creation, sequencing, evidencing, and a holistic view of context, and typically unfolds through iterative phases of research, ideation, prototyping, and implementation \cite{Stickdorn2012}.
SD has begun to shape healthcare policy by linking decision-makers with real user needs and enabling fast, testable ideas that cut through institutional barriers \cite{Mager2017}. In e-health, SD rethinks interfaces and whole systems, ensuring technology fits human behavior and supports secure, accessible, and empowering experiences \cite{Patricio2020}. One of the key SD tools for ideation is the \textit{user persona}, already applied in DT projects in sectors such as transportation, urban and environmental management, and energy.
Personas represent archetypal users, helping designers and developers better understand who they are designing for \cite{Palmer2024}.
Although SD has gained traction in healthcare innovation, service design remains absent from the current literature on digital twins, particularly those incorporating visualization, visual analytics, and HCI.
While some existing digital twin solutions implicitly adopt principles common in user-centered approach, which is one of the core principles of SD, these efforts often remain fragmented and unacknowledged.
A few studies reflect a shift in perspective from mere efficiency to user experience, as seen in works like \cite{HENNIGS202443,Schmaltz2023}.
However, even in these promising cases, the principles associated with SD are applied unconsciously, without recognizing their methodological origin or exploiting their potential.
This paper aims to highlight the value of intentionally incorporating SD principles alongside visualization and visual analytics in developing healthcare DTs, which can enhance the real-world applicability of DT solutions.

\section{State of the Art}
\label{sec:sota}

We conducted a targeted literature review to explore the current role of visualization and visual analytics in healthcare digital twins and identify whether service design principles are integrated. The review focused on papers explicitly describing digital twin systems in healthcare that include visualization, visual interfaces, interactive components, or decision support dashboards. We considered scientific contributions that model a digital twin of a patient, clinical process, or physiological aspect in healthcare, and that include visualizations, visual interfaces, and interactive components, such as dashboards, VR/AR components, or monitoring interfaces. 
We searched papers using \textit{Google Scholar}, \textit{Scopus}, \textit{DBLP}, and \textit{PubMed Central}. We also searched on the preprint archives \textit{arXiv} and \textit{medRxiv} to consider the newest research directions.
%
%
From 39 retrieved papers, we selected 20 papers after full-text review. The final corpus consists of papers published between 2023 and 2025, highlighting the recent interest in DT solutions incorporating visualizations, visual analytics, or, more generally, interactive visual interfaces. 
These 20 works span a wide range of clinical domains: 
cardiology \cite{Zhang2023,Reza2024,Li2024Toward,Qian2025,Criseo2025,Rahim2024}, rehabilitation \cite{Ha2024Hybrid, Sosa2024, Schmaltz2023}, 
diabetes \cite{Thamotharan2023,Bilal2024},
neurology \cite{Kok2025},
gastroenterology \cite{Garbey2023},
pulmonology \cite{HENNIGS202443},
sepsis \cite{Scott2024},
anatomy \cite{Choi2025},
somnology \cite{Xu2025Triboelectric},
telehealth \cite{Kim2024Assessing},
and preventive medicine \cite{Li2025Digital,Ruiu2024}.
Despite the different clinical domains and DT solutions, none of the reviewed works intentionally incorporates service design principles. Six studies \cite{Bilal2024, Criseo2025, Qian2025,Reza2024,Scott2024,Zhang2023} do not reference or apply service design concepts, while the others apply them only implicitly. 
The literature review revealed a recurring gap: a lack of explicit application of service design principles. The implicit usage of some principles suggests that visualization researchers may not recognize their connection to the broader service design discipline. One possible explanation is a lack of awareness of service design as a structured and interdisciplinary methodology.

\section{Research Agenda}
\label{sec:agenda}

Service design can help align digital twin technologies with real clinician and patient needs by designing interfaces for broader healthcare experiences, enhancing the real-world applicability of DT solutions.
To address the integration gap, we identify four main areas where SD principles can support the development of visualization-based digital twin systems in healthcare.

This research agenda outlines how SD can enhance healthcare digital twins by addressing four goals: (1) improving utility, usability, and desirability through user research and ethical reflection; (2) balancing system efficiency with meaningful user experiences; (3) designing not only interfaces but holistic user journeys that foster trust and engagement; and (4) aligning digital twin development with healthcare policies and institutional workflows.

\vspace{-2mm}
\subsection{Improving system utility, usability, and desirability} \vspace{-1mm}
\noindent 
Enhancing the effectiveness of visualization-based DT systems in healthcare requires focusing on their utility (meeting real-world needs), usability (ensuring systems are accessible and operable), and desirability (aligning with user preferences). A service design approach can guide this by coupling the design and development process with user research, visualization task modeling, and ethical reflection. 
In information visualization, it is a well-established practice to formalize user interaction intents, i.e., to articulate why users interact with visual representations \cite{Yi2007, Tominski20IVDA}. Frameworks such as the task typology by Brehmer and Munzner \cite{Brehmer2013} help structure and classify visualization tasks in terms of what users do, how they do it, and why. This task-oriented lens is essential for designing DT solutions that are not only visually rich but also functionally aligned with clinical or patient goals. User research, particularly through SD personas (archetypal representations of key user groups), helps define their characteristics and needs \cite{Palmer2024}. In healthcare DTs, personas clarify functional requirements and support customization for different user groups, such as by role, age, or diagnosis \cite{Lutze2020}. 
Visualization guidelines \cite{Paziraei2025} help select representations, but often overlook healthcare’s multi-actor nature. DT systems may also be used at the same time by a diverse group of users, including clinicians, patients, and caregivers, with distinct preferences, cognitive models, and desired levels of complexity of the DT solution. SD principles suggest designing for role-based or collaborative multi-user interaction, depending on the application context. 
Concrete use scenarios rooted in service settings further strengthen this alignment. As shown by West et al. \cite{West2024}, mapping both technical and human processes reveals opportunities for service-level innovation. 
Finally, ethical considerations must be embedded from the outset. DT systems in healthcare often collect sensitive psychological and physiological data, making them susceptible to bias, risk, and unintended consequences. Key concerns include balancing individual and collective benefits \cite{West2024}, ensuring inclusivity, mitigating algorithmic bias, and maintaining trust in the system’s outputs \cite{Tsekleves2025, Barricelli2024}. These considerations influence not only how data is handled but also how it is visualized and interpreted.

\vspace{-2mm}
\subsection{Balancing efficiency and user experience} \vspace{-1mm}
\noindent 
A DT solution should also support engaging, context-aware user experiences to be effective in real-world applications. This holds especially in sensitive domains such as rehabilitation, intensive care, or chronic condition management. To this aim, SD suggests attention to the distribution of tasks between the system and the user and the interaction modalities and situational needs.
A key starting point is user research, 
observations, interviews, and other user research methods
that uncover how users behave in their natural environments. As highlighted by Barricelli et al. \cite{Barricelli2024}, observing users can inform task allocation, i.e.,  deciding which tasks should be handled by the DT  and which should remain in the hands of the user (e.g., exploration, interpretation, decision-making, confirmation). In this way, it is possible to improve visual analytics-based digital twins, in which user interaction is essential.
Equally important in SD is the consideration of situation awareness, i.e., understanding of their environment, system status, and potential future developments.
According to Barricelli et al. \cite{Barricelli2024}, visual interfaces must be adapted to users' perceptual and cognitive needs, which vary by role, experience, and context. For instance, a clinician may require only 2D visualizations, while a patient or caregiver might benefit more from immersive VR-based feedback systems.
These choices map directly to the ``how'' dimension of Brehmer typology \cite{Brehmer2013}.

\vspace{-2mm}
\subsection{Design both user interfaces and user experience} \vspace{-1mm}
\noindent 
In healthcare DT, the interface should not be just a means of visualizing data; rather, it is a critical component of the user experience (UX) and a key determinant of system adoption, usability, and long-term engagement \cite{Shaalan2020}. SD supports visualization and interface design by emphasizing how users experience the service, including the ease with which they can learn, use, and trust a new system. 
In this context, designing for UX means designing not only what users see and click but also what they understand, feel, and expect. For instance, the development of intuitive dashboards that aim to reduce user training time and foster a smoother transition to digital workflows is valuable. For example, 
using interfaces that lower the barrier to entry and accelerate the adoption of a digital culture within the hospitals. 
From a SD perspective, these approaches reinforce the idea that the interface is just one part of a larger service journey. Good user interfaces must be embedded in a coherent and seamless experience. Moreover, valuable to this aim are frameworks of UX principles for medical interfaces \cite{Shaalan2020}.

\vspace{-2mm}
\subsection{Service and Policy Innovation} \vspace{-1mm}
\noindent 
Consideration of policy, regulation, and standards is critical to ensure the responsible development and deployment of healthcare DT solutions. As emphasized by Tsekleves et al. \cite{Tsekleves2025}, the complexity of DT technologies demands proactive alignment with healthcare regulations, ethical standards, and institutional policies. 
Service design can play a key role here by offering participatory methods that connect system designers, clinicians, and policymakers early in the innovation process. Mager \cite{Mager2017} highlights how SD brings decision-makers closer to real user needs.
By leveraging tools such as service blueprints \cite{Bitner2008} and stakeholder mapping \cite{Scholes2001}, SD supports the design of digital twin systems that align with institutional workflows and policy objectives.

\section{Conclusion}
\label{sec:conclusion}

This paper outlines an underexplored challenge in developing visualization-based healthcare digital twins: the lack of integration between service design, visualization, and visual analytics. Reviewing recent systems, we observed that, while visual and interactive components are advancing, service design principles are either absent or applied without intentionality. We argue that bridging these domains is essential to creating digital twin solutions more aligned with real-world healthcare services.


\bibliographystyle{abbrv-doi-narrow}

\bibliography{biblio}

\clearpage
\newpage

\begin{table*}[tb]
    \caption{List of the selected papers presenting digital twin systems in healthcare that include visualization, visual interfaces, interactive components, or decision support dashboard. Despite the different clinical domains and DT solutions, none of the reviewed works intentionally incorporates service design principles. Six studies do not reference or apply service design concepts, while the remaining works exhibit only implicit usage of SD principles. The list is sorted by publication year.}
    \label{tab:papers}
    \footnotesize
    \centering%
    \renewcommand{\arraystretch}{2}
    \begin{tabular}{llllc m{9.5cm}}
    \toprule
    \textbf{\#} & \textbf{Ref.} & \textbf{Year} & \textbf{Domain} & \textbf{SD Implicit Usage} & \textbf{Title}\\
    \midrule

    1 & \cite{Thamotharan2023} & 2023 & Diabetes & \sdyes & { \footnotesize Human Digital Twin for Personalized Elderly Type 2 Diabetes Management } \\\hline
    
    2 & \cite{Schmaltz2023} & 2023 & Rehabilitation & \sdyes & {\footnotesize Human Digital Twin-based interactive dashboards for informal caregivers of stroke patients} \\\hline

    3 & \cite{Zhang2023} & 2023 & Cardiology & \sdno & {\footnotesize Predicting ventricular tachycardia circuits in patients with arrhythmogenic right ventricular cardiomyopathy using genotype-specific heart digital twins} \\\hline
    
    4 & \cite{Ha2024Hybrid} & 2024 & Rehabilitation & \sdyes & {\footnotesize A Hybrid Upper-Arm-Geared Exoskeleton with AnatomicalDigital Twin for Tangible Metaverse Feedback andCommunication} \\\hline

    5 & \cite{Garbey2023} & 2024 & Gastroenterology & \sdyes & {\footnotesize Application of Digital Twin and Heuristic Computer Reasoning to  Workflow Management: Gastroenterology Outpatient Centers  Study} \\\hline

    6 & \cite{Reza2024} & 2024 & Cardiology & \sdno & {\footnotesize Assessing Post-TAVR Cardiac Conduction Abnormalities Risk Using a Digital Twin of a Beating Heart} \\\hline

    7 & \cite{Kim2024Assessing} & 2024 & Telehealth & \sdyes & {\footnotesize Holistic Patient Assessment System using Digital Twin for XR Medical Teleconsultation} \\\hline

    8 & \cite{Sosa2024} & 2024 & Rehabilitation & \sdyes & {\footnotesize Innovative Metaheuristic Optimization Approach with a Bi-Triad for Rehabilitation Exoskeletons} \\\hline

    9 & \cite{Ruiu2024} & 2024 & Preventive Medicine & \sdyes & {\footnotesize Metaverse \& Human Digital Twin: Digital Identity, Biometrics, and Privacy in the Future Virtual Worlds} \\\hline

    10 & \cite{Scott2024} & 2024 & Sepsis & \sdno & {\footnotesize Population scale proteomics enables adaptive digital twin
modelling in sepsis} \\\hline

    11 & \cite{Li2024Toward} & 2024 & Cardiology & \sdyes & {\footnotesize Toward Enabling Cardiac Digital Twins of Myocardial Infarction Using Deep
 Computational Models for Inverse Inference} \\\hline

    12 & \cite{HENNIGS202443} & 2024 & Pulmonology & \sdyes & {\footnotesize Towards a digital twin based monitoring tool for ventilated patients} \\\hline

    13 & \cite{Choi2025} & 2025 & Anatomy & \sdyes & {\footnotesize Anatomy education potential of the first digital twin of a Korean cadaver} \\\hline

    14 & \cite{Bilal2024} & 2025 & Diabetes & \sdno & {\footnotesize CitySEIRCast: an agent-based city digital twin for pandemic analysis and simulation} \\\hline

    15 & \cite{Qian2025} & 2025 & Cardiology & \sdno & {\footnotesize Developing cardiac digital twin populations powered by machine learning provides electrophysiological insights in conduction and repolarization} \\\hline

    16 & \cite{Criseo2025} & 2025 & Cardiology & \sdyes & {\footnotesize Development of a digital twin for the diagnosis of cardiac perfusion defects} \\\hline

    17 & \cite{Li2025Digital} & 2025 & Preventive Medicine & \sdyes & {\footnotesize Digital twins as global learning health and disease models for preventive and personalized medicine} \\\hline

    18 & \cite{Kok2025} & 2025 & Neurology & \sdyes & {\footnotesize MetaXAI: Metahuman-assisted audio and visual explainability framework for Internet of Medical Things} \\\hline

    19 & \cite{Rahim2024} & 2025 & Cardiology & \sdyes & {\footnotesize Personalized topology-informed localization of standard 12-lead ECG electrode placement from incomplete cardiac MRIs for efficient cardiac digital twins} \\\hline

    20 & \cite{Xu2025Triboelectric} & 2025 & Somnology & \sdyes & {\footnotesize Triboelectric Mat Multimodal Sensing System (TMMSS) Enhanced by Infrared Image Perception for Sleep and Emotion-Relevant Activity Monitoring} \\

  \bottomrule
\end{tabular}
\end{table*}


\end{document}